\documentclass[toc]{PoS}
\let\ifpdf\relax

\usepackage{amsmath}
\usepackage{amssymb,amscd}
\usepackage[arrow,matrix,curve]{xy}

\def\={\ =\ }
\def\dd{{\rm d}}

\newcommand{\tr}[1]{\:{\rm tr}\,#1}
\renewcommand{\span}{\:{\rm span}\,}

\def\slasha#1{\setbox0=\hbox{$#1$}#1\hskip-\wd0\hbox to\wd0{\hss\sl/\/\hss}}

\def\periodb#1{\setbox0=\hbox{$#1$}#1\hskip-\wd0\hbox to\wd0{-}}

\newcommand{\delder}[1]{\frac{\delta}{\delta #1}}   		

\newcommand{\bz}{{\bar{z}}}

\newcommand{\unit}{\mathbf{1}}   			

\newcommand{\CA}{\mathcal{A}}    			

\newcommand{\xd}{\dot{x}}

\newcommand{\CC}{\mathcal{C}}

\newcommand{\CG}{\mathcal{G}}

\newcommand{\CH}{\mathcal{H}}

\newcommand{\CK}{\mathcal{K}}

\newcommand{\CL}{\mathcal{L}}
\def\CrL{{%
    \setbox0\hbox{$\mathcal{L}$}%
    \rlap{\hbox to \wd0{\hss$-$\hss}}\box0
}}

\newcommand{\CO}{\mathcal{O}}

\newcommand{\CP}{\mathcal{P}}

\newcommand{\CT}{\mathcal{T}}

\newcommand{\CE}{\mathcal{E}}
\newcommand{\frg}{\mathfrak{g}}				

\newcommand{\sfs}{{\sf s}}
\newcommand{\sft}{{\sf t}}
\newcommand{\sfm}{{\sf m}}

\newcommand{\FR}{\mathbb{R}}     			
\newcommand{\FC}{\mathbb{C}}     			
\newcommand{\RZ}{\mathbb{Z}}     			
\newcommand{\CPP}{{\mathbb{C}P}}    			
\newcommand{\ah}{\hat{a}}

\newcommand{\dpar}{\partial}     			
\newcommand{\de}{\mathrm{e}}     			
\newcommand{\di}{\mathrm{i}}     			
\newcommand{\eps}{{\varepsilon}}			

\newcommand{\eand}{{\qquad\mbox{and}\qquad}}     		

\newcommand{\der}[1]{\frac{\dpar}{\dpar #1}}   		
\newcommand{\dder}[1]{\frac{\dd}{\dd #1}}   		
\newcommand{\pr}{\mathsf{pr}}     			

\newcommand{\au}{\mathfrak{u}}
\newcommand{\asu}{\mathfrak{su}}

\newcommand{\sU}{\mathsf{U}}     			

\newcommand{\sSU}{\mathsf{SU}}

\newcommand{\sLie}{\mathsf{Lie}}

\newcommand{\sEnd}{\mathsf{End}\,}

\newcommand{\remark}[1]{}     				
\def\tyng(#1){\hbox{\tiny$\yng(#1)$}}			
\def\tyoung(#1){\hbox{\tiny$\young(#1)$}}			

\newcommand{\beq}{\begin{eqnarray}}
\newcommand{\eeq}{\end{eqnarray}}

\title{Groupoid Quantization of Loop Spaces}

\ShortTitle{Groupoid Quantization of Loop Spaces \hspace{1.5in} {\rm Christian S\"amann and Richard J. Szabo}}

\author{Christian S\"amann\thanks{Speaker.} \ \ and \ Richard J. Szabo\thanks{Report numbers: \ HWM--12--02 \ , \ EMPG--12--03}%
        \\ Heriot-Watt University, Edinburgh, U.K. \\
        E-mail: \email{C.Saemann@hw.ac.uk , R.J.Szabo@hw.ac.uk}}

\abstract{We review the various contexts in which quantized 2-plectic
  manifolds are expected to appear within closed string theory and
  M-theory. We then discuss how the quantization of a 2-plectic manifold
  can be reduced to ordinary quantization of its loop space, which is a symplectic manifold. 
We demonstrate how the latter can be quantized using groupoids. After reviewing the necessary background, we present the groupoid quantization of the loop space of $\FR^3$ in some detail.}

\FullConference{Proceedings of the Corfu Summer Institute 2011 -- School and Workshops on Elementary Particle Physics and Gravity\\
		September 4--18, 2011\\
		Corfu, Greece}

\begin{document}

\section{Introduction}

A symplectic manifold $(M,\omega)$ is a manifold $M$
endowed with a closed and non-degenerate 2-form $\omega$, i.e.,
$\dd\omega=0$ and $\iota_X\omega=0$ if and only if $X=0$, where
$\iota_X$ denotes contraction with the vector field $X\in
\CC^\infty(M,TM)$. The symplectic structure induces a Poisson structure on
$M$ according to the bracket
\beq
\{f,g\}_\omega :=\iota_{X_f}\iota_{X_g}\omega \ , 
\eeq
where $X_f$
denotes the Hamiltonian vector field of a smooth function $f$ on $M$ defined by $\iota_{X_f}\omega=\dd f$. A symplectic manifold can therefore be used as a phase space in Hamiltonian mechanics. This is the starting point for geometric quantization \cite{Woodhouse:1992de}.

Multisymplectic manifolds are categorified generalizations of symplectic manifolds: A $p$-plectic manifold $(M,\varpi)$ is a manifold $M$ endowed with a closed non-degenerate $p+1$-form $\varpi$. Non-degeneracy means here again that $\iota_X\varpi=0$, $X\in \CC^\infty(M,TM)$, is equivalent to $X=0$. Note that symplectic manifolds are 1-plectic in this nomenclature: $p$ labels the level of categorification. We will be mainly interested in 2-plectic manifolds, which come with a closed non-degenerate 3-form~$\varpi$. 

A 2-plectic form $\varpi$ often induces a Nambu-Poisson 3-bracket
\cite{Nambu:1973qe,Takhtajan:1993vr} (but not always, see
\cite{springerlink:10.1007/BF00400143} for a discussion). If this is
the case, the underlying 2-plectic manifold can be used as a
multiphase space in Nambu mechanics. Many approaches to the
quantization of Nambu mechanics have been proposed over the years, see
e.g.\ the references collected in \cite{DeBellis:2010pf}. Instead of
focusing on the Nambu-Poisson structure and its quantization, we take
a 2-plectic manifold $(M,\varpi)$ as our starting point for
quantization. We show that one can map the 2-plectic structure on $M$
to a symplectic structure on the loop space of $M$ by using a
transgression map. We then quantize loop space using Hawkins' groupoid
approach to quantization \cite{Hawkins:0612363}. This article
summarizes recent work which is a
continuation of our studies in higher quantization
\cite{DeBellis:2010pf,DeBellis:2010sy,Saemann:2011zq,Saemann:2011yi}.

\section{String theory and higher quantum geometry}

There are a number of reasons why we are interested in 2-plectic
manifolds and their quantization. As a motivation for our ensuing
constructions, we start by reviewing some of the various contexts in which quantized
2-plectic manifolds are purported to appear and play a role in the
dynamics of
string theory and M-theory.

\subsection{M2-M5 brane systems and Basu-Harvey equations}

One of our main motivations stems from
the description of a configuration of M-branes that arises from a lift
of a D-brane configuration to M-theory.
Consider a stack of $k$ D1-branes extending along the $x^6$-direction and ending on a D3-brane, which extends in the $x^1,x^2,x^3$-directions:
\begin{equation}\label{diag:D1D3}
 \begin{tabular}{lccccccc}  & 0 & 1 & 2 & 3 & 4 & 5 & 6  \\
D1 & $\times$ &  & & & & & $\times$\\ D3 & $\times$ & $\times$ & $\times$ & $\times$ & 
 \end{tabular}
\end{equation}
This BPS configuration gives a string theory realisation of
magnetic monopoles \cite{Diaconescu:1996rk,Tsimpis:1998zh}: From the
perspective of the D3-branes, the endpoints of the D1-branes appear as
sources of magnetic flux and the dynamics of this system is described
by the Bogomolny monopole equations. On the other hand, we can also
study this system from the perspective of the D1-branes. Here the
system is described in terms of three $\au(k)$-valued scalar fields
$X^i$, $i=1,2 ,3$, which parametrize the fluctuations of the D1-branes along the directions parallel to the worldvolume of the D3-branes. The dynamics is governed by the Nahm equations
\begin{equation}
 \frac{\dd}{\dd x^6}X^i+\eps^{ijk}\, [X^j,X^k]=0~.
\end{equation}
A simple solution is given by $X^i=\frac{1}{x^6}\, G^i$, where
$G^i=\eps^{ijk}\, [G^j,G^k]$ generate a $k$-dimensional irreducible representation of
$\asu(2)$. This is called a fuzzy funnel \cite{Myers:1999ps}: any given point $x^6$ of the worldvolume of the D1-branes polarizes into a fuzzy sphere with radius $\frac1{x^6}$. The constant matrices $G^i$ are here interpreted as spherical coordinates on this fuzzy sphere. This polarization provides the transition between the one-dimensional worldvolume of the D1-branes and the three-dimensional worldvolume of the D3-branes.

The configuration \eqref{diag:D1D3} can be lifted to M-theory by first T-dualizing along the $x^5$-direction and subsequently interpreting $x^4$ as the M-theory direction. The resulting configuration
\begin{equation}
 \begin{tabular}{rccccccc}
 & 0 & 1 & 2 & 3 & \phantom{(}4\phantom{)} & 5 & 6 \\
M2 & $\times$ & & & & & $\times$ & $\times$ \\
M5 & $\times$ & $\times$ & $\times$ & $\times$ & $\times$ & $\times$ &
\end{tabular}
\end{equation}
yields a self-dual string \cite{Howe:1997ue} from the perspective of
the M5-brane. The perspective of the M2-branes is described by the Basu-Harvey equations~\cite{Basu:2004ed}
\begin{equation}\label{eq:BHequation}
 \frac{\dd}{\dd x^6}X^\mu+\eps^{\mu\nu\kappa\lambda}\, [X^\nu,X^\kappa,X^\lambda]=0~,
\end{equation} 
where $\mu,\nu,\ldots=1,\ldots,4$ and the scalar fields $X^\mu$ now
take values in a 3-Lie algebra: A 3-Lie algebra is an extension of the
notion of a Lie algebra to a vector space endowed with a totally
antisymmetric trilinear bracket, which also satisfies a corresponding
higher version of the Jacobi identity. Here the simple solution
corresponding to a fuzzy funnel reads as
$X^\mu=\frac{1}{\sqrt{2x^6}}\, G^\mu$, where
$G^\mu=\eps^{\mu\nu\kappa\lambda}\, [G^\nu,G^\kappa,G^\lambda]$. Clearly we would like to have an interpretation of this solution as a polarization of points of the worldvolume of the M2-branes into fuzzy three-spheres $S^3$ with radius $\frac{1}{\sqrt{2x^6}}$. Unfortunately, no consistent such interpretation exists in the literature to our knowledge. 

Just as the area form turns $S^2$ into a symplectic manifold, the
volume form of $S^3$ turns the three-sphere into a 2-plectic
manifold. A comprehensive approach to the quantization of 2-plectic
manifolds should therefore be able to fill this gap in the
interpretation of the fuzzy M2-brane funnel.

\subsection{Quantum geometry of M5-branes}

The worldvolume field theory of a flat D-brane receives
noncommutative Moyal-type deformations if a constant $B$-field
background of 10-dimensional supergravity is turned on. Consider, e.g., the background 2-form field
$B=b\,\dd x^1\wedge \dd x^2$ on a D3-brane extending in the $x^1,\ldots,x^3$-directions. This induces the
Moyal-type deformation of the coordinate algebra of the D3-brane given
by~\cite{Schomerus:1999ug,Seiberg:1999vs}
\begin{equation}
 [x^1,x^2]=\di\, \frac{\lambda^2}{1+\lambda^2} \ \stackrel{\lambda\ll 1}{\sim} \ \di\, \lambda^2~,
\label{Bfieldcomm}\end{equation}
where $\lambda=2\pi \, \alpha'\, b$. This commutation relation
describes a Heisenberg algebra which may be regarded as a quantization
of the symplectic structure on $\FR^2$ provided by the $B$-field. In~\cite{Chu:2009iv} it
is demonstrated how the Nahm equations can be used to encode boundary
conditions for open strings; the
commutation relation (\ref{Bfieldcomm}) then accounts for the induced
shift in open string boundary conditions due to a background $B$-field.

Analogously, turning on a constant 3-form $C$-field background of
11-dimensional supergravity should yield quantum deformations of the
worldvolume geometry of an M5-brane. These deformations can be
understood as accounting for the modification in boundary conditions for
open membranes ending on the M5-brane in the Basu-Harvey equations~\cite{Chu:2009iv}. For example, it was found that the constant $C$-field background 
\begin{equation}\label{eq:Cfield}
 C=\theta \, \dd x^0\wedge \dd x^1\wedge \dd x^2+\theta'\, \dd x^3\wedge \dd x^4\wedge \dd x^5
\end{equation}
gives rise to 3-Lie algebra-type deformations of the coordinate
algebra of an M5-brane extending in the $x^1,\ldots, x^5$-directions with
\begin{equation}
 [x^0,x^1,x^2]=\di\, \theta\eand [x^3,x^4,x^5]=\di\, \theta'~.
\label{3LieCfield}\end{equation}
These relations define a pair of Nambu-Heisenberg algebras. Again the $C$-field \eqref{eq:Cfield} provides a 2-plectic structure on $\FR^{1,2}\times \FR^3$, and the appropriate worldvolume deformations should be obtained from a quantization of this 2-plectic manifold.

The effect of a constant $C$-field on an M2-brane ending on an
M5-brane was also studied some time ago
in~\cite{Bergshoeff:2000jn,Kawamoto:2000zt} from the perspective of
loop space: quantization of the open membranes then yields a
noncommutative loop space structure, analogously to the way in which
quantization of open strings ending on a D-brane in a constant
$B$-field induces a noncommutative coordinate algebra. We will come back to these
results later on, thus clarifying the precise manner in which the
3-Lie algebra structure (\ref{3LieCfield}) is meant to be a
repackaging of the complicated noncommutative loop space structure~\cite{Chu:2009iv}.

\subsection{T-duality and nonassociative closed string geometry}

Quantized 2-plectic structures have also recently emerged in the
context of \emph{closed} strings.

Firstly, it has been suggested that the phase space of the
bosonic string is a 2-plectic manifold~\cite{Baez:2008bu}. A
quantization of the bosonic string would therefore correspond to a
quantization of the underlying 2-plectic manifold. Instead of
quantizing a 3-bracket, the authors of~\cite{Baez:2008bu} propose to
quantize a bracket on certain one-forms defined by
\beq
\{\alpha,\beta\}_\varpi^{(1)}:=
\iota_{X_\alpha}\,\iota_{X_\beta}\varpi
\label{1formbracket}\eeq
where $\alpha,\beta\in\Omega^1(M)$ are ``Hamiltonian one-forms'' in
the sense that there exists vector fields $X_\alpha$ and $X_\beta$ on
$M$ such that $\iota_{X_\alpha}\varpi=\dd\alpha$ and
$\iota_{X_\beta}\varpi= \dd\beta$. This bracket is skew-symmetric, but
the Jacobi identity is violated by a term proportional to
$\dd\iota_{X_\alpha}\beta$. The resulting algebra of one-forms thus
has the structure of a Lie 2-algebra. Our loop space quantization will
resolve both the conceptual problem of quantizing one-forms and the
failure of the Jacobi identity.

Secondly, closed
string 3-form flux backgrounds have recently been used to derive
nonassociative spacetime structures. For example, applying three
T-dualities to the closed string mode expansions on a three-torus
$M=T^3$ with constant 3-form $H$-flux leads to a modification of the
phase space commutation relations given by~\cite{Lust:2010iy}
\beq
[x^i,x^j]= \di\, \theta^{ijk}\,p_k \ , \qquad [x^i,p_j]= \di\, \delta_j^i \eand
[p_i,p_j]= 0 \ .
\label{Lustalg}\eeq
This bracket defines a pre-Lie algebra structure but not a Lie
algebra, as the
Jacobi identity is not satisfied: the Jacobiator defines a nonassociative 3-bracket structure
$[x^i,x^j,x^k]= \di \, \theta^{ijk}$. It should also be properly understood
in a categorified framework as a (semi-strict) Lie 2-algebra. Such
nonassociative structures have been confirmed through an analysis of
closed string three-point functions in the $\sSU(2)$ WZW
model~\cite{Blumenhagen:2010hj}, and in an $H$-linearized expansion of conformal field
theory on flat space~\cite{Blumenhagen:2011ph} in both geometric and
non-geometric 3-form flux backgrounds. It raises the
tantalizing possibility of deriving a theory of noncommutative and nonassociative gravity from non-geometric flux
  compactifications.

The appearence of nonassociativity in $H$-flux backgrounds is not new
and was observed some time ago in the context of open string theory~\cite{Cornalba:2001sm},
where it was found that open string two-point functions reproduce the
noncommutative and nonassociative Kontsevich star-product for twisted
Poisson structures. We can understand the loss of associativity in the
following way. Just like the commutator is a quantization of a Poisson
bracket which encodes the failure of a product to
be commutative, the quantization of a Nambu-Poisson 3-bracket should
encode the failure of a product to be associative. This suggests that
appropriately quantized 2-plectic manifolds do not give rise to
ordinary Hilbert spaces and linear operator algebras on them, but to
some more general structures. In the following we will describe an
attempt to understand these structures in the context of loop space
quantization; this is very natural from the point of view of closed strings.

That there does still exist a suitable $C^*$-algebraic framework can
be seen by understanding the relation to \emph{topological}
nonassociative
tori~\cite{Bouwknegt:2004ap,Ellwood:2006my,Grange:2006es}. The general
setting is where the target space is a principal torus bundle
\beq
M \ \xrightarrow{ \ T^d \ } \ W 
\eeq
of rank $d$ with quantized three-form $H$-flux $[H] \in H^3(M,\RZ)$. This
background and its duals are neatly described in an algebraic
language: There is a continuous trace $C^*$-algebra $\CA$ with
Dixmier-Douady class $[H]$ whose spectrum is $X$. The T-dual algebra
which arises
from dualizing along the fibre directions is then a crossed-product $\widehat\CA= \CA\rtimes_\alpha
\widehat\FR{}^d$ of $\CA$ with the Pontrjagin dual of the abelian
group $\FR^d$ along the action $\alpha:\FR^d\rightarrow {\sf
  Aut}(\CA)$ induced by translations in the fibres. The algebra
$\widehat\CA$ is not a continuous trace algebra in general and so does not
have a well-defined geometric spectrum; hence it gives a ``global''
description of some of the recent non-geometric
backgrounds which have appeared in string theory.

To elucidate this description, let us consider the simplest case where
$M=T^3$ is a trivial $T^d$-bundle over $W=T^{3-d}$. For $d=0$ this is
just the three-torus $T^3$, which we endow with $H$-flux and $B$-field
given by
\beq
H= k\, \dd
x^1\wedge \dd x^2\wedge \dd x^3= \dd B \eand B= k\, x^1\, \dd
x^2\wedge\dd x^3 \ ,~~~k\in\RZ~.
\eeq
In the doubled geometry the nonassociativity is manifested in the dual
coordinates $\tilde x^i$ as~\cite{Lust:2010iy,Blumenhagen:2011ph}
\beq\label{eq:WindingAlgebra}
[\tilde x^2,\tilde x^3] \sim w_1 \eand [\tilde x^1,\tilde
x^2,\tilde x^3] \neq 0~,
\eeq
where $w_i$ are winding modes. We now describe each of the three
T-duals in turn:
\begin{itemize}
\item For $d=1$, T-dualising along the $x^3$-direction yields a
  continuous trace $C^*$-algebra whose spectrum is the Heisenberg
  nilmanifold, i.e., the quotient of the three-dimensional Heisenberg
  group $H_\FR$ by a cocompact lattice $H_\RZ$. The nonassociativity structure becomes
\beq
[\tilde x^2,x^3] \sim w_1 \eand [\tilde x^1,\tilde
x^2,x^3] \neq 0 \ .
\eeq
\item For $d=2$, performing an additional T-duality transformation
  along the $x^2$-direction yields the
  dual algebra $\widehat\CA= C^*(H_\RZ)\otimes\CK$, where $\CK$ is
  the algebra of compact operators and $C^*(H_\RZ)$ is the convolution
  $C^*$-algebra of the lattice $H_\RZ$ regarded as an abelian
  group. This algebra describes a fibration of noncommutative two-tori
  $T_\theta^2$ over $S^1$ with deformation parameter $\theta= k\,
    x^1$ induced by the $B$-field and varying over the base. This is
    the global description of a \emph{T-fold}, i.e., a locally
    Riemannian space whose transition functions between coordinate
    patches involve T-duality transformations. The noncommutativity
    along the fibre directions appears in the linearized brackets
\beq
[x^2, x^3]  \sim w_1 \eand [\tilde x^1,
x^2,x^3] \neq 0 \ .
\eeq
\item For $d=3$, the additional T-duality along $x^1$ is strictly
  speaking forbidden as the $B$-field background is not invariant
  under translations around this direction. Nevertheless,
  considerations from string field theory suggest that the resulting
  background should be included when considering the totality of
  closed string backgrounds. Unlike the T-fold, this
  background does not admit even a local description as a Riemannian
  manifold, and is called a non-geometric $R$-flux
  background. Algebraically, the dual algebra $\widehat\CA=\CK\big(L^2(\,\widehat
  T{}^3\,)\big)\rtimes_{u_\phi}\widehat T{}^3$ is called the nonassociative three-torus
  $T_\phi^3$. It is described by a convolution product which is
  twisted by a 3-cocycle $\phi\in Z^3\big(\,\widehat T{}^3,\sU(1)\big)$
  associated to the 3-form $H$. The linearized nonassociativity
  structure is encoded in the brackets
\beq
[x^2, x^3] \sim p_1 \eand [x^1,
x^2,x^3] \neq  0 \ .
\eeq
This nonassociative algebra may be regarded as a $C^*$-algebra in a
monoidal category~\cite{Bouwknegt:0702802}; it may thus be identified as the
convolution $C^*$-algebra of the Lie 2-group which integrates the Lie
2-algebra (\ref{Lustalg}).
\end{itemize}

\section{Quantization of symplectic manifolds}

We will now review two well-known quantization prescriptions, which
are both combined in the groupoid approach to
quantization. First, we recall geometric and Berezin-Toeplitz
quantization using the example of the complex projective line
$\CPP^1$. Second, we review the quantization of the dual of a Lie
algebra using the example of $\kappa$-Minkowski space. 

\subsection{Berezin-Toeplitz quantization}

The construction of the Hilbert space in Berezin-Toeplitz quantization is identical to that of geometric quantization. It is only in the construction of the quantization maps that the two prescriptions differ.

We call a symplectic manifold $(M,\omega)$ {\em quantizable} if there
is a {\em prequantum line bundle} over $M$. A prequantum line bundle
$(L,h,\nabla)$ consist of a line bundle $L\to M$ together with a hermitian metric $h$ and a connection $\nabla$ that is compatible with $h$. Moreover, we demand that the {\em quantization condition} 
\begin{equation}\label{eq:quantcondition}
 F_\nabla:=\nabla^2 =-2\pi\, \di\, \omega
\end{equation}
is satisfied, which implies that $[\omega] \in H^2(M,\RZ)$. Note that
a positive tensor power of a prequantum line bundle is again a
prequantum line bundle. The datum $(L,h,\nabla)$ is called a {\em
  prequantization} of the symplectic manifold $(M,\omega)$. The
quantization condition \eqref{eq:quantcondition} implies that $L$ is
positive or ample. Therefore, the space of global sections of $L$ is
interesting enough to be identified with the Hilbert space $\CH$ in
our quantization. It is well-known, however, that taking smooth
sections of $L$ yields a Hilbert space which is too large. To restrict
the Hilbert space to the appropriate size, one introduces a {\em
  polarization}, i.e.,\ a smooth integrable lagrangian
distribution. We can then define the Hilbert space as the subspace of
smooth global sections that are annihilated by the polarization. For
us, it will be convenient to work with K\"ahler polarization, i.e., we regard $M$ as a complex manifold and use the anti-holomorphic tangent bundle as a distribution. The resulting Hilbert space is thus given 
by global holomorphic sections of the prequantum line bundle $L$.

In the case $(\CPP^1,\omega)$, where $\omega$ is the usual K\"ahler
form giving rise to the Fubini-Study metric on $\CPP^1$, we can use
the positive tensor powers of the hyperplane line bundle $\CO(1)$ (the
dual of the tautological line bundle) as prequantum line bundles,
i.e., we define $L_k:=\CO(k)$. The finite-dimensional Hilbert space
$\CH_k:=H^0(\CPP^1,L_k)$ can then be identified with the space of homogeneous polynomials of degree $k$ in the homogeneous coordinates $z_\alpha$, $\alpha=1,2$, on $\CPP^1$. This space is isomorphic to the $k$-particle Fock space in the Hilbert space of two harmonic oscillators with creation and annihilation operators $\ah^\dagger_\alpha$ and $\ah_\alpha$ satisfying $[\ah_\alpha,\ah_\beta^\dagger]=\delta_{\alpha\beta}$ and $\ah_\alpha|0\rangle=0$:
\begin{equation}
 \CH_k\cong\span_\FC(z_{\alpha_1}\cdots z_{\alpha_k})=\span_\FC
 \big(\ah^\dagger_{\alpha_1}\cdots \ah^\dagger_{\alpha_k}|0\rangle \big)~.
\end{equation}

To construct a map between functions on $M$ and endomorphisms on $\CH$, we use the Rawnsley coherent states. They form an overcomplete set of states labelled by points on $M$, and they can be constructed explicitly for any quantizable symplectic manifold. In the case of $\CPP^1$, they are the truncated Glauber coherent states
\begin{equation}
 |z\rangle =\tfrac{1}{k!}\, \big(\bar{z}_\alpha \, \ah^\dagger_\alpha\big)^k|0\rangle~.
\end{equation}
From these, we construct the {\em coherent state projector}
\begin{equation}
 \CP(z):=\frac{|z\rangle\langle z|}{\langle z|z\rangle}~,
\end{equation}
which is simultaneously an endomorphism on the Hilbert space $\CH$ and a smooth function on $M$. It therefore provides a bridge between the quantum and classical pictures, and we can use it to define quantization maps.

First, let us introduce the injective Berezin symbol
\begin{equation}
 \sigma\,:\,  \sEnd(\CH) \ \longrightarrow \ \Sigma\subset
 \CC^\infty(M)~, \qquad \sigma(\hat{f}\, )(z)=\tr_\CH\big(\CP(z) \, \hat{f}\, )~.
\end{equation}
The smooth functions in $\Sigma:= \sigma(\sEnd(\CH))$ form the set of quantizable functions, and we define a quantization map
\begin{equation}
 \widehat{-}\, : \, \Sigma \ \longrightarrow \ \sEnd(\CH)~, \qquad \hat{f}=\sigma^{-1}(f)~.
\end{equation}
Alternatively, we can introduce the Berezin-Toeplitz quantization map
\begin{equation}
 T\, :\, \CC^\infty(M) \ \longrightarrow \ \sEnd(\CH)~, \qquad T(f):=\int_M\, \dd \mu(z)~ \CP(z)\, f(z)~,
\end{equation}
where $\dd\mu(z)$ is the canonical Liouville measure on $M$.

In the case of the Riemann sphere $\CPP^1$, the functions in $\Sigma_k=\sigma(\sEnd(\CH_k))$ are of the form\footnote{The function space $\Sigma_k$ is the linear span of the spherical harmonics $Y_{\ell m}$ on $S^2$ with $\ell\leq k$, $|m|\leq \ell$.} 
\begin{equation}
 f(z)=\sum_{\alpha_i,\beta_i=1,2} \, f^{\alpha_1\cdots
   \alpha_k\beta_1\cdots \beta_k} \, \frac{z_{\alpha_1}\cdots
   z_{\alpha_k}\, \bz_{\beta_1}\cdots \bz_{\beta_k}}{|z|^{2k}}~,
\end{equation}
where $f^{\alpha_1\cdots \alpha_k\beta_1\cdots
  \beta_k}=(f^{\beta_1\cdots \beta_k\alpha_1\cdots
  \alpha_k})^*\in\FC$, $|z|^2:= \bz_\alpha\, z_\alpha$, 
and the quantization map reads explicitly as 
\begin{equation}
 f(z) \ \longmapsto \ \hat{f}=\sum_{\alpha_i,\beta_i=1,2} \,
 f^{\alpha_1\cdots \alpha_k\beta_1\cdots
   \beta_k}\, \frac{1}{k!} \, \ah^\dagger_{\alpha_1}\cdots
 \ah^\dagger_{\alpha_k}|0\rangle\langle 0|\ah_{\beta_1}\cdots \ah_{\beta_k}~.
\end{equation}
The quantization map $\widehat{-}$ therefore creates normal ordered operators. The quantization map $T$ gives the corresponding anti-normal ordered operators. For further details on Berezin-Toeplitz quantization, see e.g.\ \cite{IuliuLazaroiu:2008pk} and references therein.

\subsection{Quantization of the dual of a Lie algebra}

The groupoid approach to quantization is an extension of the well-known procedure of quantizing the dual of a Lie algebra, which we now briefly review.

We start from a Lie group $G$ with Lie algebra $\frg$. There is a
natural Poisson structure on $\frg^*$ that is defined as follows:
Linear functions on $\frg^*$ can be identified with elements of $\frg$,
and for those functions we define $\{g_1,g_2\}(x):=\langle x,[g_1,g_2]\rangle$,
where $g_1,g_2\in \frg$, $x\in\frg^*$ and $\langle
-,-\rangle:\frg^*\times\frg\to \FC$ denotes the dual pairing. Via the Leibniz identity, this Poisson bracket extends to polynomial functions on $\frg^*$, which in turn are dense in $\CC^\infty(\frg^*)$. 

To quantize the resulting Poisson algebra, we perform the following
steps~\cite{JSTOR:2374874}: We first Fourier transform to obtain elements in
$\CC^\infty(\frg)$. We then identify $\frg$ with $G$ in a neighborhood
of the identity element via the exponential mapping. On $G$, we can
use the convolution product between functions induced by the group
multiplication and the Baker-Campbell-Hausdorff formula. We can then perform the inverse operations to pullback
the result to $\CC^\infty(\frg^*)$. For nilpotent Lie algebras, the exponential map between $\frg$ and $G$ is a global diffeomorphism. In this case, the above construction is equivalent to both Kontsevich's deformation quantization and quantization via the universal enveloping algebra of $\frg$~\cite{Kathotia}.

To illustrate the procedure, let us use it to quantize $d$-dimensional
$\kappa$-Minkowski space, which we identify with the dual $\frg^*$ of
a Lie algebra $\frg$. The generators $g^0,g^1,\dots,g^{d-1}$ of $\frg$
are naturally identified as coordinate functions on $\frg^*$, and they
have the Lie brackets
\begin{equation}
 [g^0,g^i]=\frac{\di}{\kappa}~g^i \eand [g^i,g^j]=0 \qquad \mbox{for}
 \quad 0<i,j<d
\end{equation}
with a constant $\kappa>0$. The corresponding Lie group $G$ is
generated by $W(k_0,\vec k\, )=V_{\vec k}\, U_{k_0}$ with
$U_{k_0}:=\exp\big(\di\, k_0\, g^0\big)$ and  $V_{\vec
  k}:= \exp\big(-\di\, \mbox{$\sum_i$}\, k_i\, g^i\big)$. The resulting group multiplication reads as
\begin{equation}
 W(k_0,\vec k\,)\, W(k'_0,\vec k'\,) = W(k_0+k_0',\vec k+\de^{-
  k_0/\kappa}\, \vec k'\,)~.
\end{equation}
We start by Fourier transforming a function $f\in \CC^\infty(\frg^*)$ via
\begin{equation}
 \tilde f(g)= \int_{\frg^*}~ \dd \mu_{\frg^*}(x)\ \de^{-2\pi\, \di\, \langle
  x,g\rangle}\, f(x)~,
\end{equation}
where $\dd \mu_{\frg^*}(x)$ denotes the invariant Haar measure on
$\frg^*$ regarded as an additive abelian group. We then identify $\frg$ with $G$ via the exponential mapping, and define the inverse Fourier transform according to 
\begin{equation}
 W(\tilde f\, ):= \int_G~ \dd\mu_G(p_0,\vec{p}\, )~ W(p_0,\vec{p}\, )
 \, \tilde f(p_0,\vec{p}\, )~,
\end{equation}
where $\dd\mu_G(p_0,\vec{p}\, )=\de^{p_0/ \kappa}\, \dd p_0\, \dd
\vec{p}$ is the invariant Haar measure on $G$. The convolution product
$\circledast$ of two functions is defined via $W(\tilde{f}\,
) \, W(\tilde{f}'\, )=W(\tilde{f}\circledast\tilde{f}'\, )$, and here we obtain
\begin{equation}
 (\tilde f\circledast \tilde f'\,)(p_0,\vec p\, )= \int_{G}~
\dd p_0' \ \dd \vec p\,' \ \de^{ p_0'/ \kappa} \ \tilde f(p_0',\vec p\,'\,)\,
\tilde f\big(p_0-p_0'\,,\,\de^{
  p_0'/ \kappa}\, (\vec p-\vec p\,'\, )\big)~.
\end{equation}

In a similar way one readily quantizes any Poisson vector space
$(\FR^d,\pi)$ with constant Poisson bivector $\pi$, which yields the usual direct sums of Moyal planes and ordinary vector spaces.

\section{Groupoid approach to quantization}

\subsection{From groups to groupoids}

Any group may be regarded as a small category with one object $\unit$, in which all morphisms, which correspond to the group elements, are invertible. Dropping the restriction that there is only one unit, we arrive at a groupoid: A {\em groupoid} is a small category in which every morphism is invertible. Analogously to Lie groups, we introduce {\em Lie groupoids}: Here both sets of objects and morphisms form manifolds and all morphisms are smooth.

Equivalently, we can describe a groupoid by a set $B$ of units (called
the base) and a set $\CG$ of morphisms or arrows between units (also called the groupoid) together with the following maps:
\begin{itemize}
 \item The source and target maps $\sfs,\sft:\CG\rightrightarrows B$ that yield the tail and the head of an arrow.
 \item The object inclusion map $\unit: B\hookrightarrow \CG$, which yields an arrow starting and ending at the given unit and therefore satisfies $\sfs(\unit(x))=\sft(\unit(x))=x$ for all $x\in B$.
 \item An associative partial multiplication $\sfm:\CG\times \CG\to \CG$ that concatenates
   two composable arrows. It is thus defined for arrows $(g,h)\in
   \CG\times \CG$ with $\sft(g)=\sfs(h)$ (the ``2-nerve'' of $\CG$), and we have $\sfs(\sfm(g,h))=\sfs(g)$ and $\sft(\sfm(g,h))=\sft(h)$, as well as $\sfm(g,\unit(\sft(g)))=g$ and $\sfm(\unit(\sfs(g)),g)=g$.
\end{itemize}
Finally, we demand that each arrow $g\in\CG$ has a two-sided inverse
$g^{-1}$ corresponding to the inverse arrow, i.e.,
$\sfs(g^{-1})=\sft(g)$, $\sft(g^{-1})=\sfs(g)$,
$\sfm(g,g^{-1})=\unit(\sfs(g))$, and $\sfm(g^{-1},g)=\unit(\sft(g))$. In a
Lie groupoid, the sets $B$ and $\CG$ are manifolds and the groupoid maps $\sfs,\sft,\unit,\sfm$ are all smooth.

A trivial example of a groupoid is a group $G$, where $B$ consists of
a single point, $\CG$ is identified with $G$, the source and target maps are
trivial, and the inclusion map of the point yields the identity element of $G$. A
non-trivial example that will reappear in slightly altered form later
on is that of the {\em pair groupoid}. Given a manifold $M$, we use
points in $M$ as units, $B=M$, and consider all possible arrows having
their heads and tails in $M$, $\CG=M\times M$. The identity map is
thus $\unit(x)=(x,x)$ and the other structure maps of the groupoid are defined by
\begin{equation}
 x \ \xrightarrow{~~(x,y)~~} \ y \ \xrightarrow{~~(y,z)~~} \ z
\end{equation}
for all $x,y,z\in M$.

The linearized version of a Lie group is a Lie algebra, which is given
by the tangent space to the Lie group at the identity element. Because
a Lie groupoid has more than one unit, one has to consider the union
of all these tangent spaces. This gives rise to the Lie algebroid
$\sLie(\CG)$ of a Lie groupoid $\CG$ which is a vector bundle over $B$
with total space
\begin{equation}
 \sLie(\CG)=\bigcup_{x\in B}\, T_{\unit(x)}\big(\sft^{-1}(x)\big) \
 \subset \ T\CG~.
\end{equation}

More generally, a {\em Lie algebroid} $\CE$ is a vector bundle $\CE$
over a base manifold $B$ endowed with a Lie bracket $[-,-]_\CE$ on
smooth sections of $\CE$ and a bundle morphism $\rho:\CE\rightarrow
TB$, called the {\em anchor map}, which is compatible with the Lie
bracket on sections, i.e., the tangent map to $\rho$ is a Lie algebra homomorphism,
\begin{equation}
\dd\rho_{[\psi_1,{\psi_2}]_\CE} =[\dd \rho_{\psi_1},\dd
\rho_{\psi_2}]_{TB}~, \qquad
   {\psi_1},{\psi_2}\in\CC^\infty(B,\CE)~,
\end{equation}
and a Leibniz rule is satisfied when multiplying sections of $\CE$ by smooth functions on $B$,
\begin{equation}\label{eq:LALeibniz}
[{\psi_1},f\,
{\psi_2}]_\CE=f\,[{\psi_1},{\psi_2}]_\CE+\rho_{\psi_1}(f)\,
{\psi_2}~,\qquad {\psi_1},{\psi_2}\in\CC^\infty(B,\CE)~, \qquad f\in \CC^\infty(B)~.
\end{equation}
We thus have the fibrations
\begin{equation}
\xymatrix{
 \CE \ \ar[r]^{\rho} \ar[dr] & \ TB \ar[d] \\
   & \ B
}
\end{equation}
The Lie algebroid of a Lie groupoid which is also a Lie group is just
the usual Lie algebra together with the trivial anchor map. In the
case of the pair groupoid $\CG=M\times M$, the corresponding Lie algebroid is the
tangent bundle
\begin{equation}
 \sLie(M\times M)=\bigcup_{x\in M} \, x\times T_xM=TM~.
\end{equation}

Above we derived Lie algebroids from Lie groupoids. While the inverse operation is defined for Lie algebras, which we can integrate to Lie groups, the same is not true in general for Lie algebroids. We will encounter a criterion for integrability in the next subsection. For more details on groupoids and algebroids, see e.g.\ \cite{0521499283}.

\subsection{Summary of groupoid quantization}

The starting point of any quantization is a Poisson manifold
$(M,\pi)$, where $\pi$ is the bivector field that encodes the Poisson
bracket through $\{f,g\}_\pi:=\pi(\dd f,\dd g)$ for all
$f,g\in\CC^\infty(M)$. A Poisson manifold naturally comes with a Lie
algebroid structure on its cotangent bundle $T^*M$: The anchor map
$\rho:T^*M\rightarrow TM$ is defined as $\alpha\mapsto \pi(\alpha, -
)$. The compatible Lie bracket on exact one-forms is given by $[\dd
f,\dd g]_{T^*M} =\dd \{f,g\}_\pi$, which can be extended via the Leibniz rule \eqref{eq:LALeibniz} to arbitrary one-forms. Explicitly, we have
\begin{equation}
 \big\langle [\alpha,\beta]_{T^*M} \,,\,X \big\rangle= \big\langle
 \alpha\,,\,[\pi,\langle\beta,X\rangle]_{\rm S} \big\rangle-
 \big\langle \beta\,,\,[\pi,\langle\alpha,X\rangle]_{\rm S}
 \big\rangle-[\pi,X]_{\rm S}(\alpha,\beta)~,
\end{equation}
where $\alpha,\beta\in\Omega^1(M)$, $X$ is a smooth vector field and
$[-,-]_{\rm S}$ on the right-hand side is the Schouten-Nijenhuis bracket of multivector fields on $M$.

When quantizing the dual of a Lie algebra, we used the convolution
algebra on the integrating Lie group. It is thus natural to expect
that we can define a quantization of (the dual of) a Lie algebroid by
using the convolution algebra on the integrating Lie groupoid. The
fact that not all Lie algebroids can be integrated merely corresponds
to the fact that not all Poisson manifolds can be quantized with this
method. The idea of using Lie groupoids in quantization goes back to
work by Karas\"ev, Weinstein and Zakrzewski, see
\cite{Hawkins:0612363} and references therein. Bonechi \emph{et al.}
have recently addressed the problem of quantizing $\CPP^1$ to the Podle\`s sphere with this method \cite{Bonechi:2010yh}.

A subtle issue in quantization using groupoids is the introduction of
a polarization. Here we will follow the approach proposed by Hawkins
\cite{Hawkins:0612363}, which we now briefly describe.

We start from a Poisson manifold $(M,\pi)$, which gives rise to the
corresponding Lie algebroid $T^*M$ as described above. Instead of
integrating $T^*M$ as a Lie algebroid, we will find an {\em
  integrating symplectic Lie groupoid} for $M$, i.e., we construct a
Lie groupoid $\Sigma$ with base $M$, where $\Sigma$ is a symplectic
manifold with symplectic structure $\omega$. We demand that $\sft$ is
a Poisson map\footnote{In some treatments it is also demanded that
  $\sfs$ is an anti-Poisson map. We do not impose this restriction here.} and that $\omega$ is multiplicative. It can be shown \cite{Crainic:2002aa} that the existence of an integrating symplectic Lie groupoid for $M$ is equivalent to the integrability 
of the Lie algebroid $T^*M$. Let us briefly spell out the definition
of multiplicative forms on $\Sigma$: Consider the set of composable
arrows $\Sigma_{[2]}\subset\Sigma\times \Sigma$ in $\Sigma$. We then have projection maps $\pr_{1},\pr_{2}:\Sigma_{[2]}\rightarrow \Sigma$ onto the first and second arrow in the composable pair:
\begin{equation}
 \pr_1(g,h)=g \qquad \mbox{and} \qquad \pr_2(g,h)=h~.
\end{equation}
A form $\alpha$ on $\Sigma$ is called {\em multiplicative} if
\begin{equation}
 \pr_1^*\alpha+\pr_2^*\alpha=\sfm^*\alpha~.
\end{equation}

Having found an integrating symplectic groupoid for the Poisson
manifold we wish to quantize, we prequantize $\Sigma$ as a symplectic
manifold in the second step, i.e., we construct a prequantum line
bundle $(L,h,\nabla)$ over $\Sigma$ such that $F_\nabla=-2\pi\, \di\, \omega$. 

Third, we have to endow $\Sigma$ with a groupoid polarization, i.e., we have to introduce a smooth integrable lagrangian distribution that reduces the sections of $L$ over $\Sigma$. In our examples, a groupoid polarization is either chosen to be given by K\"ahler polarization or it is derived from the fact that the base $M$ is a lagrangian submanifold in the integrating symplectic Lie groupoid $\Sigma$.

In the final step, we construct a polarized convolution algebra of
$\Sigma$, using e.g.~ a Haar system of measures, which is twisted by a $\sU(1)$-valued 2-cocycle. In the
cases where there is a globally defined symplectic potential on
$\Sigma$, the cocycle twist essentially encodes the failure of the
potential to be multiplicative; in general its role is to induce a
multiplication on the fibres of $L\to\Sigma$. This twisted polarized convolution algebra is then identified with the algebra of functions on the quantization of~$M$.

It should be stressed that there are existence and uniqueness issues remaining open in many of the steps of this quantization procedure. One advantage of this approach is that it avoids the explicit construction of the usual Hilbert space, yielding the quantized algebra of functions rather directly. This fact might help generally in the quantization of 2-plectic manifolds, where we expect non-associative structures which cannot be modeled on linear spaces.

\subsection{Groupoid quantization of $\FR^2$}

The simplest example of a groupoid quantization is that of the real
vector space $M=\FR^2$
with Poisson structure given by a constant bivector field
$\pi=\pi^{ij}\, \der{x^i}\wedge \der{x^j}$, $i,j=1,2$. As integrating
Lie groupoid, we choose the cotangent bundle $\Sigma=T^*M =M\times M^*$, which we coordinatize by $(x^i,p_i)$. The symplectic structure on $\Sigma$ is the natural one, $\omega=\dd x^i\wedge \dd p_i$. The groupoid maps are given by the diagram
\begin{equation}
 x^i+\tfrac{1}{2}\, \pi^{ij}\, p_j~~\xrightarrow{~~(x^i,p_i)~~}~~
 x^i-\tfrac{1}{2}\, \pi^{ij}\, p_j~,
\end{equation}
so that we have
\begin{equation}
 \sfs(x^i,p_i)=(x^i+\tfrac{1}{2}\,\pi^{ij}\, p_j)\eand
 \sft(x^i,p_i)=(x^i-\tfrac{1}{2}\,\pi^{ij}\,p_j)~,
\end{equation}
and we recognize the Bopp shifts familiar from canonical quantization of $\FR^2$. One readily checks that $\sft$ is indeed a Poisson map, i.e.,\
\begin{equation}
 \{\sft^*f,\sft^*g\}_\omega =\sft^*\{f,g\}_\pi \ .
\end{equation}
It remains to verify the multiplicativity of $\omega$. Consider the composition of arrows
\begin{equation}
 x^i+\tfrac{1}{2}\, \pi^{ij}\,
 (p_j+p'_j)~\xrightarrow{(x^i+\tfrac{1}{2}\,\pi^{ij}\,p_j,p'_i\,
   )}~x^i+\tfrac{1}{2}\, \pi^{ij}\,
 (p_j-p'_j)~\xrightarrow{(x^i-\tfrac{1}{2}\,\pi^{ij}\,p_j',p_i)}~x^i-\tfrac{1}{2}\,
 \pi^{ij}\, (p_j+p'_j)~.
\end{equation}
Therefore, $\Sigma_{[2]}$ can be identified with $M\times M^*\times
M^*$, which we coordinatize by $(x^i,p_i,p'_i\, )$ and we have the structure maps
\begin{equation}
\begin{aligned}
 \pr_1(x^i,p_i,p'_i\, )=(x^i+\tfrac{1}{2}\,\pi^{ij}\,p_j,p'_i\,
 )&~,~~~\pr_2(x^i,p_i,p'_i\,
 )=(x^i-\tfrac{1}{2}\,\pi^{ij}\,p_j',p_i)~,\\ \mbox{and} \quad
 \sfm(x^i,p_i,p'_i\, )&=(x^i,p_i+p'_i\, )~.
\end{aligned}
\end{equation}
It is easy to verify that $\pr_1^*\omega+\pr_2^*\omega=\sfm^*\omega$, and the groupoid $\Sigma$ is thus indeed an integrating symplectic groupoid for $M$.

The prequantization of $(\Sigma,\omega)$ is now straightforward: Since
$\omega$ is exact, we take $L$ to be the trivial line bundle over
$\Sigma$ with the obvious hermitian metric $h$ and connection $\nabla$
such that $F_\nabla=-2\pi\, \di\, \omega$.

A globally defined symplectic potential for $\omega$ is given by
$\vartheta=-x^i\, \dd p_i$, and as polarization we choose the vector fields in $T\Sigma$ that are in the kernel of $\vartheta$. This reduces functions on $\Sigma$ to functions on $M^*$. The twist element $\sigma_\pi$ encodes the failure of $\vartheta$ to be multiplicative, and we have
\begin{equation}
-\di\, \sigma_\pi^{-1}\, \dd\sigma_\pi:=
(\pr^*_1+\pr_2^*-\sfm^*)\vartheta=\dd\big(-\tfrac{1}{2}\, p_i \,
\pi^{ij}\, p_j'\, \big)~,
\end{equation}
which yields
\begin{equation}
\sigma_\pi(p,p'\,)=\de^{-\frac{\di}{2}\, p_i\, \pi^{ij}\, p_j'\,
 } ~.
\end{equation}
Together with the natural translation-invariant measure on $M^*$, we
thus obtain the twisted convolution product on polarized functions on
$\Sigma$ given by
\begin{equation}
 (\tilde f\circledast_\pi \tilde g\, )(p)=\int_{M^*}~ \dd p' \
 \sigma_\pi(p',p-p'\, )\ \tilde f(p'\, )\, \tilde g(p-p'\,)~,
\end{equation}
which is the usual Moyal product after Fourier transformation to momentum space.

The groupoid quantization of the two-dimensional torus $T^2$ is
completely analogous, except that one now has to invoke a
Bohr-Sommerfeld quantization condition on the leaves of the
polarization on the cotangent groupoid $\Sigma\cong T^2\times \FR^2$,
which reproduces the usual deformation of Fourier series on $T^2$~\cite{Hawkins:0612363}.

Analogously, one can quantize $\kappa$-Minkowski space, where the
integrating groupoid is the cotangent bundle $\Sigma=T^*G$, with $G$ the Lie group of the
$\kappa$-Minkowski Lie algebra; the cocycle twist here is
trivial. Berezin-Toeplitz quantization of K\"ahler manifolds also fits
into this framework; the integrating groupoid here is the pair groupoid, and again the cocycle twist is trivial.

\section{Quantization of loop spaces}

\subsection{2-plectic manifolds, gerbes and loop spaces}

As discussed above, a symplectic manifold $(M,\omega)$ with symplectic
form $\omega$ that satisfies the quantization condition $[\omega] \in
H^2(M,\RZ)$ comes naturally with a prequantum line bundle, i.e., there
is a line bundle $L\to M$ with connection $\nabla$ such that
$F_{\nabla}=-2\pi\, \di\, \omega$. Analogously, a 2-plectic manifold
$(M,\varpi)$ with 2-plectic form $\varpi$ satisfying $[\varpi] \in
H^3(M,\RZ)$ comes naturally with a prequantum abelian gerbe, i.e.,\ an
abelian gerbe with 2-connection whose curvature is $H=-2\pi\, \di\, \varpi$. 

Since a gerbe may be thought of as a categorification of a vector bundle, an obvious approach to the quantization of 2-plectic manifolds is to categorify conventional quantization, see e.g.~\cite{Baez:2009:aa,Rogers:2010sc}. We will report on progress in categorifying Hawkins' groupoid approach in future work.

Here we will follow a different approach: We can map the prequantum
gerbe on the 2-plectic manifold $(M,\varpi)$ to a prequantum line
bundle on the loop space\footnote{When refering to loop space, we
  will usually mean loops modulo reparametrization invariance.} of $M$
using a {\em transgression map}. Consider the double fibration in
which the correspondence space $\CL M\times S^1$ is simultaneously
fibred over a manifold $M$ and its free loop space $\CL M:=\CC^\infty(S^1,M)$:
\begin{equation}
\xymatrix{
 & \CL M\times S^1 \ar[dl]_{ev} \ar[dr]^{pr} & \\
M &  & \CL M
} 
\end{equation}
The projection $pr$ is the obvious one, and the map $ev$ is the
evaluation $x(\tau)$ of a loop $x\in \CL M$ at a given angle $\tau\in
S^1$. We can now construct a map $\CT:\Omega^{n+1}(M)\rightarrow
\Omega^n(\CL M)$ by pulling back an $n+1$-form $\alpha$ on $M$ along
$ev$ and projecting it down to $\CL M$ by integrating over the circle
fibres. The latter map amounts to filling up one slot of the form with
the tangent vector to the loop. Explicitly, the transgression of a
form $\alpha=\frac{1}{(n+1)!}\, \alpha_{\mu_1\ldots\mu_{n+1}}\, \dd
x^{\mu_1}\wedge\cdots\wedge \dd x^{\mu_{n+1}}$ is given by
\begin{equation}
 (\CT\alpha)(x):=\big((pr_!\circ ev^*)\alpha \big)(x)=\oint\, \dd
 \tau~ \frac{1}{n!}\, \alpha_{\mu_1\ldots\mu_n\mu_{n+1}} \big(x(\tau)\big)~\xd^{\mu_{n+1}}(\tau)~\delta x^{\mu_1}(\tau)\wedge \cdots \wedge \delta x^{\mu_n}(\tau)~,
\end{equation}
where $\xd^\mu(\tau)=\dder{\tau}x^\mu(\tau)$ denotes the tangent
vector to the loop at the angle $\tau$. The transgression map is a
chain map: It maps closed forms to closed forms and exact forms to
exact forms. There is therefore a well-defined restriction
$\CT:H^{n+1}(M,\RZ)\rightarrow H^n(\CL M,\RZ)$. More generally, for $\alpha\in
\Omega^{n+1}(M)$, $n\geq 0$, one has $\delta \CT\alpha=\CT \dd\alpha$,
where $\dd=\dd x^\mu \, \der{x^\mu}$ and $\delta=\oint\, \dd \tau~
\delta x^{\mu}(\tau)\, \delder{x^\mu(\tau)}$ denote the exterior
derivatives on $M$ and $\CL M$, respectively; in particular, $\CT\,
\dd f=0$ for all $f\in\CC^\infty(M)$. The transgression of a
form is also invariant under reparametrizations of the loop. The map
$\CT$ is not surjective, and there are more line bundles over $\CL M$
than gerbes on $M$. The inverse map, called ``regression'', is
generally defined only on the image of $\CT$.

This point of view has been successfully used in lifting the ADHMN construction of magnetic monopoles to a construction of self-dual strings in M-theory \cite{Saemann:2010cp,Palmer:2011vx}.

Let us consider the explicit example $M= \FR^3$ with 2-plectic form
$\varpi=\frac1{3!}\, \eps_{ijk}\, \dd x^i\wedge \dd x^j\wedge \dd
x^k$. Transgressing this form yields a symplectic 2-form on the loop
space $\CL\FR^3$ given by
\begin{equation}
\omega=\CT\varpi= \oint\, \dd \tau \ \frac12\, \eps_{ijk}~\xd^k(\tau)~\delta x^i(\tau)\wedge \delta x^j(\tau)~.
\end{equation}
Note that the kernel of $\omega$ is non-trivial, as
\begin{equation}
 \iota_{X_\alpha} (\CT\varpi)=0 \qquad \mbox{for}~~~X_\alpha=\oint\,
 \dd \rho \ \alpha(\rho)~\xd^i(\rho)\, \delder{x^i(\rho)}~.
\end{equation}
The vector field $X_{\alpha}$, however, generates reparametrization
transformations and it is not contained in the space of derivations
acting on functions of reparametrization invariant loops in $\CL
\FR^3$. Restricting to such loops, we can therefore either invert the components of $\omega$  or, equivalently, follow the usual construction via Hamiltonian vector fields to arrive at the Poisson bracket
\begin{equation}
 \{f,g\}_\omega :=\oint\, \dd \tau \ \oint\, \dd \rho \
 \delta(\tau-\rho) \, \pi^{ijk}\, \frac{\xd_k(\rho)}{\big|\xd(\rho)
   \big|^2}\, \Big(\, \delder{x^i(\tau)}f\Big)\, \Big(\, \delder{x^j(\rho)}g\Big)
\end{equation}
on $\CC^\infty(\CL \FR^3)$. This expression for the Poisson bracket is reparametrization invariant.
  
\subsection{Groupoid quantization of $\CL\FR^3$}

We now come to the quantization of the loop space $\CL M$ for
$M=\FR^3$ using the groupoid approach. Up to a few subtleties, we can
follow closely the discussion of the case $\FR^2$. As integrating
groupoid, we take $\Sigma=T^*\CL M\cong\CL T^*M$. There are two points
about this identification that we have to stress. First of all, we
exclude elements $\alpha=\oint\, \dd \tau~\alpha_i(\tau)~\delta
x^i(\tau)$ of $T^* \CL M$ that have coefficients $\alpha_i(\tau)$
which are distributional. Second, the identification
is only local and does not respect reparametrization invariance. Thus,
if we coordinatize elements of $\CL T^* M$ by pairs
$(x^i(\tau),p_i(\tau))$, then the momenta $p_i(\tau)$ are not
invariant under reparametrizations of the loop but transform like the
velocity vectors $\xd^i(\tau)$. This will be visible in all formulas below.

Comparison with the case $\FR^2$ suggests introducing the source and target maps
\begin{equation}
 \sfs\big(x^{i}(\tau)\,,\,p_{i}(\tau)\big)=x^{i}(\tau) +\tfrac{1}{2}\,
 \pi^{ijk}\, p_{j}(\tau) \, \frac{\xd_{k}(\tau)}{\big|\xd(\tau)
   \big|^2}~,~~~~~~~\sft\big(x^{i}(\tau) \,,\,p_{i}(\tau)
 \big)=x^{i}(\tau)-\tfrac{1}{2}\, \pi^{ijk}\, p_{j}(\tau) \,
 \frac{\xd_{k}(\tau)}{\big|\xd(\tau) \big|^2}~,
\end{equation}
which are local in the loop parameter and reparametrization
invariant. The set of composable arrows is $\Sigma_{[2]}=\CL(M\times
M^* \times M^* )$, and the projections and multiplication maps are
\begin{equation}
\begin{aligned}
 \pr_1\big(x^{i}(\tau),p_{i}(\tau),p'_{i}(\tau) \big)&:=\big(x^{i}(\tau)+\tfrac{1}{2}\,
   \pi^{ijk}\, p_{j}(\tau)\, \frac{\xd_{k}(\tau)}{\big|\xd(\tau) \big|^2}\,,\,p'_{i}(\tau)\big)~,~~~\\[4pt]
 \pr_2\big(x^{i}(\tau),p_{i}(\tau),p'_{i}(\tau)
 \big)&:=\big(x^{i}(\tau)-\tfrac{1}{2}\, \pi^{ijk}\, p'_{j}(\tau)\,
 \frac{\xd_{k}(\tau)}{\big|\xd(\tau)\big|^2}\,,\, p_{i}(\tau)\big)~,~~~\\[4pt]
 \sfm\big(x^{i}(\tau),p_{i}(\tau),p'_{i}(\tau)\big)&:=\big(x^{i}(\tau)\,,\,p_{i}(\tau)+p'_{i}(\tau)\big)~.  
\end{aligned}
\end{equation}
The natural symplectic structure on $\Sigma$ reads as
\begin{equation}
 \omega=\oint\, \dd \tau\ \oint\, \dd \rho\ \delta(\tau-\rho)\, \delta x^{i}(\tau)\wedge \delta p_{i}(\rho)~,
\end{equation}
and $\omega$ can be derived from the symplectic potential
\begin{equation}\label{eq:vartheta}
 \vartheta=\oint \dd \rho\ x^{i}(\rho)~\delta p_{i}(\rho)~.
\end{equation}
Note that both loop space forms $\omega$ and $\vartheta$ are reparametrization invariant.

It turns out that the target map $\sft$ is a Poisson map only to lowest order in $\pi$, i.e.,\
\begin{equation}
\begin{aligned}
 \{\sft^*f,\sft^*g\}_\omega:=& \ \oint\, \dd\tau \ \bigg( \Big(\,
 \delder{x^{i}(\tau)}\sft^*f\Big)\, \Big(\,
 \delder{p_{i}(\tau)}\sft^*g\Big)-\Big(\,
 \delder{x^{i}(\tau)}\sft^*g\Big)\,
 \Big(\delder{p_{i}(\tau)}\sft^*f\Big) \bigg) \\[4pt]
=& \ \sft^*\{f,g\}_\pi+\CO(\pi^2) \qquad \mbox{for} \quad f,g\in \CC^\infty(\CL \FR^3)~.
\end{aligned}
\end{equation}
Similarly, the symplectic form $\omega$ is multiplicative only to lowest order in $\pi$: $(\pr_1^*+\pr_2^*-\sfm^*)\omega=\CO(\pi)$. Higher corrections to the groupoid structure can be computed order by order in $\pi$. Here we will just develop the quantization to lowest order.

We now follow the usual remaining steps of the groupoid approach to
quantization. To prequantize $(\Sigma,\omega)$, we introduce a trivial
line bundle $L$ over $\Sigma$ together with a connection $\nabla$ such
that $F_\nabla=-2\pi\, \di\, \omega=-2\pi\, \di\, \CT\varpi$. The
polarization is again induced by the symplectic potential, i.e., we
consider a distribution in the kernel of $\vartheta$ given by
\eqref{eq:vartheta}. This polarization renders functions independent
of $x(\tau)$, but note that it preserves the dependence on
$\xd(\tau)$. The latter point is related to the problem of introducing polarizations on 2-plectic manifolds. 

Following the algorithm, we now introduce a twist element from the failure of $\vartheta$ to be a multiplicative one-form, 
\begin{equation}
 (\pr_1^*+\pr_2^*-\sfm^*)\vartheta=-\di\, \sigma_\pi^{-1}~\delta\sigma_\pi~,
\end{equation}
and obtain
\begin{equation}
 \sigma_\pi\big(p_i(\rho),p'_j(\tau) \big)=\exp\big( -\tfrac{\di}{2}\,
 p_i(\rho)~\pi^{ijk}~\xd_k(\tau)~p_j'(\tau)~\delta(\rho-\tau) \big)
 ~.
\end{equation}
It is difficult to introduce the notion of a twisted polarized
convolution algebra on the groupoid $\Sigma$, as it is not clear how
to define a reparametrization invariant system of Haar measures on
$\Sigma$. However, by analogy with the case $\FR^2$, we conclude that
the quantization of $\Sigma$ yields a coordinate algebra on loop space
that is deformed as
\begin{equation}\label{eq:LoopCommutator}
 {}\big[\hat{x}^{i}(\tau)\,,\,\hat{x}^{j}(\rho)\big]=\tfrac{\di}{2}\,
 \pi^{ijk}\, \widehat{\frac{\xd_{k}(\tau)}{\big|\xd(\tau)\big|^2}} \ \delta(\tau-\rho)~.
\end{equation}

The expression \eqref{eq:LoopCommutator} agrees via transgression with
a quantization of one-forms as suggested by the Poisson-like brackets
developed in \cite{Baez:2008bu}. In general, with $\omega=\CT\varpi$,
the transgression of (\ref{1formbracket}) yields
\beq
\CT\{\alpha,\beta\}^{(1)}_\varpi=\{\CT\alpha,\CT\beta\}_\omega \ ,
\eeq
and so maps the problem of quantizing one-forms on $M$ to quantizing
functions on loop space. Since
$\CT\dd\iota_{X_\alpha}\beta=0$, the Jacobi identity is recovered.
Moreover, it agrees with results of
M-theory computations presented in
\cite{Bergshoeff:2000jn,Kawamoto:2000zt} using Dirac quantization of
open membrane dynamics on an M5-brane. In
\cite{Kawamoto:2000zt,Berman:2004jv}, corrections to
\eqref{eq:LoopCommutator} in higher orders of $\pi$ have been
computed. It might be interesting to compare these corrections to the
corrections which arise from adjusting the groupoid structure to higher orders in $\pi$.

Again, an analogous groupoid quantization can be applied to the loop
space of the three-dimensional torus $T^3$, with interesting
topological features similar to those arising in closed string
sigma-models with $T^3$ target space. In
particular, the algebra (\ref{eq:LoopCommutator}) agrees with L\"ust's
algebra (\ref{Lustalg}) via the identification $p_i\sim \dot x_i$. Moreover, 
the algebra \eqref{eq:LoopCommutator} agrees essentially with \eqref{eq:WindingAlgebra}
after integrating over $\tau$ and $\rho$.

\section{Conclusions and outlook}

In this article we reviewed Hawkins' groupoid approach to
quantization. This approach is particularly suited for higher
quantization, as it circumvents the introduction of the usual Hilbert
space. We demonstrated how the groupoid approach can be applied to the
quantization of 2-plectic manifolds via the detour through loop
spaces: The prequantum gerbe on a 2-plectic manifold yields a
prequantum line bundle on the corresponding loop space, which can be
subjected to ordinary quantization. In making the transition to loop
space, we traded the difficulties in dealing with gerbes for the
problems of working with infinite-dimensional manifolds. Moreover, the
expected nonassociativity in the gerbe picture has been resolved
through associative structures acting on non-separable Hilbert
spaces. The deformed coordinate algebra on loop space that we found
via this method agrees with earlier M-theory results and anticipated
closed string nonassociativity.

A number of interesting open questions, however, still remain: First
of all, one should examine if it is possible to follow the other, more
natural approach to the quantization of 2-plectic manifolds by
categorifying the groupoid approach. Second, it would be useful to
study the ensuing higher bracket structures and compare them to those
that have already been studied in the literature, see e.g.\
\cite{deAzcarraga:2010mr} and references therein. Finally, once a satisfying quantization of
$S^3$ is obtained, it would be interesting to use the resulting
quantum algebra in the Basu-Harvey equation \eqref{eq:BHequation} and
ultimately in the various M2-brane models that have been studied over
the last four years. We expect that this would contribute to the
solution of some of the as yet open problems of these models.

\section*{Acknowledgments}

This work was supported in part by the
Consolidated Grant ST/J000310/1 from the UK Science and Technology Facilities Council. The
work of CS was supported by a Career Acceleration Fellowship from the
UK Engineering and Physical Sciences Research Council.

\providecommand{\href}[2]{#2}\begingroup\raggedright\endgroup
 
\end{document}